\documentclass{aa}  
\usepackage{graphicx}
\usepackage{txfonts}
\usepackage{hyperref}
\usepackage{graphicx}
\usepackage{amsmath}
\usepackage{soul}
\usepackage{xcolor}
\usepackage{enumitem} 

\newcommand{\G}{\textit{Gaia}}
\newcommand{\V}{\texttt{VOSA}}

\newcommand{\WDs}{white dwarfs}

\usepackage{lscape}
\defcitealias{2023MNRAS.518.5106J}{JE+23}
\defcitealias{RM+19}{RM+19}

\begin{document} 
   \title{\G{} white dwarfs with infrared excess}

   \subtitle{I. The 100\,pc catalogue}

    \author{R. Murillo-Ojeda\inst{1}
    \and F. M. Jim\'enez-Esteban\inst{1} 
    \and A. Rebassa-Mansergas\inst{2,3}
    \and S. Torres\inst{2,3}}  

   \institute{Centro de Astrobiolog\'{\i}a (CAB), CSIC-INTA, Camino Bajo del Castillo s/n, 28692, Villanueva de la Ca\~nada, Madrid, Spain\\
   \email{rmurillo@cab.inta-csic.es}
   \and
   Departament de F\'{\i}sica, Universitat Polit\`{e}cnica de Catalunya, c/Esteve Terrades 5, 08860 Castelldefels, Spain
   \and
    Institut d'Estudis Espacials de Catalunya, c/Esteve Terradas 1, Edifici RDIT, Campus PMT-UPC, 08860 Castelldefels, Spain}
   \date{Received October 15, 2025; accepted Month DD, YYYY}
 
  \abstract
   {The presence of infrared excess flux observed in white dwarfs is related to the existence of debris disks or substellar companions. These systems provide important clues in the study of extrasolar planetary material and binary evolution. However, fully characterising their properties requires a statistically significant, complete sample.
   }
   {This work aims to identify a complete sample of white dwarfs with infrared excess emission within 100\,pc of the Sun. 
   }
   {We built the spectral energy distributions (SEDs) of the white dwarfs using synthetic photometry in 56 optical filters of the Javalambre Physics of the Accelerating Universe Astronomical Survey system, generated from \G{} Data Release 3 low-resolution spectra and complemented with the latest infrared photometry available at the Virtual Observatory (VO). \texttt{VOSA} was used to fit the SEDs with different atmospheric white dwarf models depending on the source spectral type. We visually checked optical and infrared images to identify contaminated photometry.
   }
   {We built a catalogue of 456 infrared excess white dwarfs, of which 292 are robust identifications, and 164 are candidates. 
   351 ($\sim$75\%) are new identifications. This implies a fraction of infrared excess white dwarfs between 5.9$\pm$0.3\% and $9.2\pm$0.4\%, higher than previous works, but in agreement with some more recent estimates.
   Furthermore, for the sample of infrared excess white dwarfs, the fraction of sources with non-hydrogen atmosphere increases with the \G{} $G_\mathrm{BP}-G_\mathrm{RP}$ colour, contrary to the general white dwarf population.
   However, this result should be interpreted with caution. 
   Additionally, a thorough comparison of our catalogue with those of previous studies was performed.
   }
  {The sample of white dwarfs with infrared excess emission within 100\,pc presented in this work is the largest, most complete and reliable to date. Due to their proximity, they are ideal targets for follow-up studies aimed at characterising circumstellar disks, substellar companions, and the composition of accreted planetary material.
  }


   \maketitle

\section{Introduction}already
\label{sec:intro}

The vast majority of stars will become, or have already become, white dwarfs \citep[e.g.][]{2010A&ARv..18..471A}. White dwarfs are therefore very common in our Galaxy and carry important information about their progenitors. They address key problems in astrophysics, serving as estimators of Galactic evolution and star formation history \citep{Rebassa2015, Torres2021, Lam2025}, reliable cosmic chronometers \citep{Fontaineetal2001, Rebassaetal2023}, and natural laboratories for testing fundamental physics under extreme conditions \citep{Isern2022}. Moreover, white dwarfs have been identified as key objects for probing the elemental composition of extrasolar planetary material through the study of metal abundances in their atmospheres \citep{Jura2003}.

Due to their high density, heavier elements than hydrogen diffuse to the deep interiors of white dwarfs. As a consequence, a large percentage of them ($\sim$70\%; \citealt{Torreseral2023, 2023MNRAS.518.5106J, Manseretal2024}) show hydrogen lines in their spectra. However, helium or carbon can arise to the surface during the evolution -- that is, cooling -- of white dwarfs, especially when they cool enough to become convective \citep{Koester+Kepler2019, Rollandetal2020, Bedard2024}. Moreover, 25-50\% of them show metal-pollution in their atmospheres as revealed by optical and/or ultraviolet spectroscopy \citep{Zuckermanetal2003, Zuckermanetal2010, Koesteretal2014}, which is thought to arise from accretion of planetary material \citep{Jura2003, Jura+Young2014}. 
In this scenario, a surviving planet (although a brown dwarf can not be ruled out; see, e.g., \citealt{Casewell+24}) perturbs the orbit of a minor body (asteroid-, comet-, or moon-like in composition). The minor body enters the white dwarf tidal radius, becomes tidally disrupted, and its fragments are eventually accreted onto the white dwarf  \citep{Debes+Sigurdsson2002, Mustilletal2024, Verasetal2024}.

The disruption of a minor body is expected to form a circumstellar dust or gas disk around the white dwarf \citep{Debesetal2012, Verasetal2014, Malamud+Perets2020a, Malamud+Perets2020b}, from which material is accreted onto its surface \citep{Verasetal2015, Malamudetal2021, Lietal2021, Brouwersetal2022}. There is overwhelming observational evidence for the existence of both dust and gas disks around white dwarfs, supporting the above scenario. Dust disks have been detected at infrared wavelengths for about 1-4\% of all white dwarfs \footnote{Note that $\lesssim 0.5$\% of white dwarfs with infrared excess are expected to be due to the presence of a brown dwarf companion.} \citep[e.g.][]{Rochettoetal2015, Farihi2016, RM+19, Wilsonetal2019, Rogersetal2020, Wang23}.
Gas disks are considerably less common, but the typical double-peaked spectral features have been observed in the optical for a few dozen white dwarfs \citep[e.g.][]{Gaensickeetal2006, Farihietal2012, Brinkworthetal2012, Debesetal2012b, Melisetal2020, Owens+23}.
Moreover, the existence of circumstellar material is corroborated by the observation of transits caused by disintegrating bodies \citep[e.g.][]{Vanderburgetal2015, Manseretal2019, Guidryetal2021, Farihietal2022, Aungwerojwitetal2024}. 
Additional evidence for the ongoing accretion process in metal-polluted white dwarfs arises from the detection of X-rays \citep{Cunninghametal2022}. Once accretion ensues on the white dwarf, the diffusion times are expected to range from $\sim $yr in hydrogen-rich atmospheres to $\sim$Myr in helium-rich ones \citep{Harrisonetal2021}. 
The diffusion timescales of helium-rich atmospheres are presumably longer than the typical lifetime of dusty disks \citep{Girvenetal2012, Cunninghametal2021}. This may explain why the percentage of white dwarfs that display infrared excess due to the existence of an accretion disk is considerably smaller than the fraction showing metal-pollution in their atmospheres. Another possibility is that these disks are too faint to be detected.

Infrared excess white dwarfs are excellent candidates for spectroscopic follow-up studies aimed at revealing the chemical composition of the accreted bodies and hence of extra-solar planetary material \citep[e.g.][]{Farihietal2013, Raddietal2015, Xuetal2017, Hollandsetal2018, Swanetal2019, Badenas-Agustietal2024}.
However, most of the existing infrared excess white dwarf samples are magnitude-limited, which prevents deriving realistic determinations of the incidence of metal-polluted white dwarfs as a function of effective temperature and mass. In this paper, we aimed at solving this issue via making use of the \G{} Data Release 3 \citep[DR3;][]{GaiaDR32023} to search for new infrared excess white dwarfs among all objects located within 100\,pc from the Sun. 

The 100\,pc white dwarf sample is expected to be a nearly-complete volume-limited sample \citep{JE+18}. The work presented here superseded the one we performed in \citet{RM+19} (hereafter \citetalias{RM+19}), which made use of the former second Data Release of \G{}. Now, \G{} DR3 provides low-resolution spectra which allowed building more robust optical spectral energy distributions (SEDs) as compared to those obtained in \citetalias{RM+19}. Moreover, a larger number of white dwarfs now have available infrared photometry from different surveys, which considerably increased the initial sample of study. In addition, with the aim of presenting a sample of infrared excess white dwarfs as pure as possible, we performed a thorough search for contaminants. Finally, we carried out an exhaustive comparison between our catalogue and those in the literature.
In a second publication, we will perform stellar parameter determination and analysis of the full catalogue, including also a discussion on the origin of the detected excess (arising either from a circumstellar disk or from a low-mass/brown dwarf companion).

The paper is organised as follows. In Section\,\ref{sec:meth} we introduce our methodology for identifying infrared excess white dwarfs. In Section\,\ref{sec:Results} we describe our results. In Section\,\ref{sec:Comparisons} we compare our catalogue with previously published samples. In Section\,\ref{sec:fract} we discussed the infrared excess fraction obtained, and we conclude the work in Section\,\ref{sec:Conclusions}.

\section{Identification of sources with infrared excess emission}
\label{sec:meth}

We identified the infrared excess from the analysis of the SEDs, as in \citetalias{RM+19}. 

\subsection{The 100 pc sample}
\label{100pc}

The initial sample of this work contained 8150 white dwarfs spectroscopically classified and characterised by \cite{2023MNRAS.518.5106J} (hereafter \citetalias{2023MNRAS.518.5106J}). In that work, the authors built a volume-limited catalogue of 12\,718 white dwarfs within 100\,pc from the Sun.
The white dwarf candidates of \G{} DR3 were selected from their position in the \G{} Hertzsprung-Russell diagram after imposing the following astrometric and photometric quality criteria:

\begin{enumerate}[label=(\roman*)]
  \item $\omega-3\sigma_{\omega}\ge 10$\,mas and $\omega/\sigma_{\omega}\ge10$; where $\omega$ is the parallax and $\sigma_{\omega}$ is its error.
  \item $F_{\rm BP}/\sigma_{F_{\rm BP}}\ge10$ and $F_{\rm RP}/\sigma_{F_{\rm RP}}\ge10$; where $F_{\rm BP}$ and $F_{\rm RP}$ are the integrated BP and RP mean fluxes,  and $\sigma_{F_{\rm BP}}$ and $\sigma_{F_{\rm RP}}$ are their respective errors.
  \item RUWE < 1.4, where RUWE stands for Renormalised Unit Weight Error; values of RUWE > 1.4 indicates a poor astrometric solution \citep{Lindegren+21}.
  \item $|C^{*}|<3\sigma_{C^{*}}$; where $C^{*}$ is the corrected BP and RP flux excess factor and $\sigma_{C^{*}}$ its scatter following the prescription by \cite{Riello+21}.
\end{enumerate}

Additionally, the authors fitted the SEDs of 8150 white dwarfs with $G_{BP}-G_{RP}<0.86$ ($T_\mathrm{eff}\gtrsim5500\,K$) with models to classify them as either pure hydrogen atmosphere (DA) white dwarfs or non-DA white dwarfs, the latter lacking hydrogen and represented in that work by helium atmospheres.
We refer the reader to \citetalias{2023MNRAS.518.5106J} for a detailed explanation of the classification methodology. These 8150 spectroscopically classified objects were our initial search sample for infrared excess emission.

\subsection{Spectral Energy Distributions}

We built the SEDs from the optical to the mid-infrared wavelength range. For the optical, we used GaiaXPy\footnote{\url{https://gaia-dpci.github.io/GaiaXPy-website/}} to obtain synthetic photometry from the \G{} DR3 spectra using all their coefficients \citep{2021A&A...652A..86C}. For the infrared ($\lambda \geq 12\,000$ \AA), we compiled photometric data from public surveys in the Virtual Observatory (VO\footnote{The VO (\url{https://www.ivoa.net/about/what-is-vo.html}) is an initiative of the International Virtual Observatory Alliance (IVOA, \url{https://www.ivoa.net/}). It was created to optimise the mining of astronomical data. The VO provides services and tools that follow the FAIR principles: findable, accessible, interoperable, and reusable.}) with the aid of \texttt{Topcat} \citep{Taylor05}. 

\subsubsection{Optical photometry}
\label{OptPhot}

We used the same methodology as in \citetalias{2023MNRAS.518.5106J} to obtain the optical photometry from the \G{} DR3 spectra. Thus, we selected the photometric system of Javalambre Physics of the Accelerating Universe Astronomical Survey (J-PAS\footnote{\url{http://www.j-pas.org/}}, \citealt{benitez2014jpas, Marin-Franch12}) for the synthetic photometry obtained from the \G{} spectra. This selection was made for the following two reasons:
i) it samples practically the whole wavelength range covered by \G{} ($\approx$\,3300$-$10\,500\,\AA) with a total of 60 filters: 54 narrow filters ($\approx$\,3780$-$9100\,\AA, with a full width at half-maximum $\approx$\,145\,\AA), 2 wider filters in the reddest and bluest part of the wavelength range, and 4 SDSS-like filters; 
ii) the spectral resolution of the photo-spectra obtained is similar to \G{} DR3 low-resolution spectra ($R \approx 60$).
Due to the low signal-to-noise ratio at the bluest end of the \G{} DR3 spectra \citep{Montegriffoetal2023}, we only kept J-PAS filters with effective $\lambda \geq 4000$ \AA, reducing the number of J-PAS filters to 56. We also imposed a quality cut in photometry, keeping only data with flux relative errors lower than $10\,\%$, and only selecting objects with at least four J-PAS photometry points.
All 8150 initial objects met these criteria.

\subsubsection{Infrared photometry}
\label{sec:IRphot}

To extend the coverage of the SED to longer wavelengths ($\lambda >12 \,000$ \AA ), we searched for available photometry at the following surveys:
the Two Micron All Sky Survey (2MASS, \citealt{2006AJ....131.1163S}), the UKIRT Infrared Deep Sky Survey (UKIDSS, \citealt{10.1111/j.1365-2966.2005.09969.x}), the Visible and Infrared Survey Telescope for Astronomy (VISTA, \citealt{refId0}), the Wide-field Infrared Survey Explorer (WISE, \citealt{Wright_2010}), Spitzer \citep{Werner_2004}, the Midcourse Space Experiment (MSX, \citealt{1996AJ....112.2862E}), AKARI \citep{2007PASJ...59S.369M}, and the Infrared Astronomical Satellite (IRAS, \citealt{iras}).
Note that the Y and Z VISTA and UKIDSS photometry were excluded because this region of the SED was already covered by the synthetic J-PAS photometry.

In order to avoid misidentifications of high proper motion white dwarfs, we translated the \G{} coordinates to an epoch closer to the mean observation epoch of each survey using the proper motion provided by \G{} DR3. 
We used epoch J2000 for 2MASS, MSX and IRAS, J2005 and J2010 for AKARI and Spitzer, respectively, J2015.4 for WISE, and finally, J2016 for UKIDSS and VISTA. 
We then limited the search into a $3\arcsec$ radius. 
The search for counterparts in AKARI, IRAS, and MSX returned no results for our white dwarf sample.

Afterwards, we cleaned up the photometry found by selecting those data with good quality flags. 
Table \ref{tab:IR_phot_table} shows filters and quality flags used for each survey. It also includes a note listing the used catalogues.

\begin{table}[ht]
    \caption[]{Infrared data from VO archives used to build the SEDs.}
    \label{tab:IR_phot_table}
    \centering
    \small
        \begin{tabular}{c c c c}
        \noalign{\smallskip}
        \hline
        \hline
        \noalign{\smallskip}
        Survey & Facility/Instrument (Filters)$^{(c)}$ & Quality flags\\
        \noalign{\smallskip}
        \hline
        \noalign{\smallskip}
        2MASS & 2MASS (J, H, K$_{s}$) & Qflg $\in (A,B,C,D)$\\
        \noalign{\smallskip}
        UKIDSS & UKIRT/WFCAM (J, H, K) & $\left|ppErrBits\right| < 256$\\
        \noalign{\smallskip}
        VISTA & Paranal/VIRCAM (J, H, K$_{s}$) & $\left|ppErrBits\right| < 256$\\
        \noalign{\smallskip}
        WISE & WISE (W1, W2) & ---\\
        \noalign{\smallskip}
        Spitzer & \begin{tabular}{c} IRAC (I1, I2, I3, I4)\\ MIPS (24\,$\mu$m, 70\,$\mu$m, 160\,$\mu$m) \end{tabular} & $^{(a)}q = 0$ or $^{(b)}n = 0$\\ 
        \noalign{\smallskip}
        \hline
        \noalign{\smallskip}
        \end{tabular}
        \tablefoot{\small{The catalogues used from each survey were: i) 2MASS: 2MASS All-Sky Catalog of Point Sources; ii) UKIDSS: Large Area Survey (LAS) DR10, Galactic Clusters Survey (GCS) DR10, Galactic Plane Survey (GPS) DR7, Deep Extragalactic Survey (DXS) DR10; iii) VISTA: VISTA Variables in the Via Lactea (VVV) DR5, VISTA Kilo-Degree Infrared Galaxy (VIKING) DR4, VISTA Hemisphere Survey (VHS) DR6, VISTA Magellanic Survey (VMS) DR4; iv) WISE: CatWISE2020 \citep{2021ApJS..253....8M}; v) Spitzer: $^{(a)}$Galactic Legacy Infrared Midplane Survey Extraordinaire (GLIMPSE) I, II $\&$ 3D, $^{(b)}$The Spitzer Enhanced Imaging Products (SEIP) source list.\\ $^{(c)}$Detailed information on the mentioned filters can be found at the \texttt{SVO Filter Profile Service "Carlos Rodrigo"}}\footnote{\url{http://svo2.cab.inta-csic.es/theory/fps/}} \citep{Rodrigo24}.}
\end{table}

In order to have reliable infrared excess detection, we only keep SEDs with at least three photometric points at wavelengths longer than 12\,000\,\AA. This resulted in a sample of 5084 objects to be searched for infrared excess emission.

Thus, the analysed SEDs contained at least seven photometric points --four in the optical and three in the infrared-- up to a maximum of 71, with $\sim$70\% of them containing 59 or more photometric points.

\subsection{Excess emission identification}
\label{sec:VOSAfit}

We used the \texttt{Virtual Observatory SED Analyzer} (\V{}\footnote{\url{http://svo2.cab.inta-csic.es/theory/vosa}}, \citealt{2008_VOSA}), to identify possible infrared excess emission by single object model SED fitting.

\V\ tool performed two automatic analyses of the photometric data to detect any possible flux excess emission. The first analysis consisted of calculating the regression slope ($\alpha$, \citealt{Lada_2006}) of the observational data in the $\log{\nu F_{\nu}}$ vs. $\log\nu$ flux-frequency plane, starting at the first point with $\lambda>21\,500$ \AA\, and adding one new data point at a time in increasing wavelength direction. If the regression slope, $\alpha$, taking into account its error, was smaller than 2.56, then \V{} flagged the data point as affected by infrared excess. The flagged data points were not longer considered in the subsequent model fitting.

After this first analysis, we fitted the SEDs with white dwarf atmospheric models of \citet{2010MmSAI..81..921K}. We used DA and non-DA models depending on the spectral classification by \citetalias{2023MNRAS.518.5106J} (see Sect. \ref{100pc}).
This spectral classification had an accuracy of over 90\% when using the Montreal White Dwarf Database (MWDD) classification as reference, and fairly agrees with a later classification based on more sophisticated artificial intelligence algorithms \citep{Garcia-Zamora2025}. We refer the reader to \citetalias{2023MNRAS.518.5106J} for a more detailed description of the grid of models. Then, \V{} performed the second automatic analysis, consisting of comparing the observational photometry with the synthetic photometry obtained from the model that best fits the data, that is, the fit with the lowest value of the reduced chi-squared parameter ($\chi^{2}$). Thus, if i) the observed flux was significantly larger ($>20\,\%$) than the predicted one, and ii) the observed flux was above the predicted one by more than 3 times the observational error ($F_{obs}-F_{mod} > 3 \sigma_{F_{obs}}$), then the photometric point was marked as affected by excess emission. Additionally, based on a visual inspection of all the fits, we manually modified for a few cases the first filter affected by infrared excess at $\lambda<21\,500\,$\AA.

\begin{figure*}[ht]
    \sidecaption
    \centering
    \includegraphics[width=0.35 \linewidth]{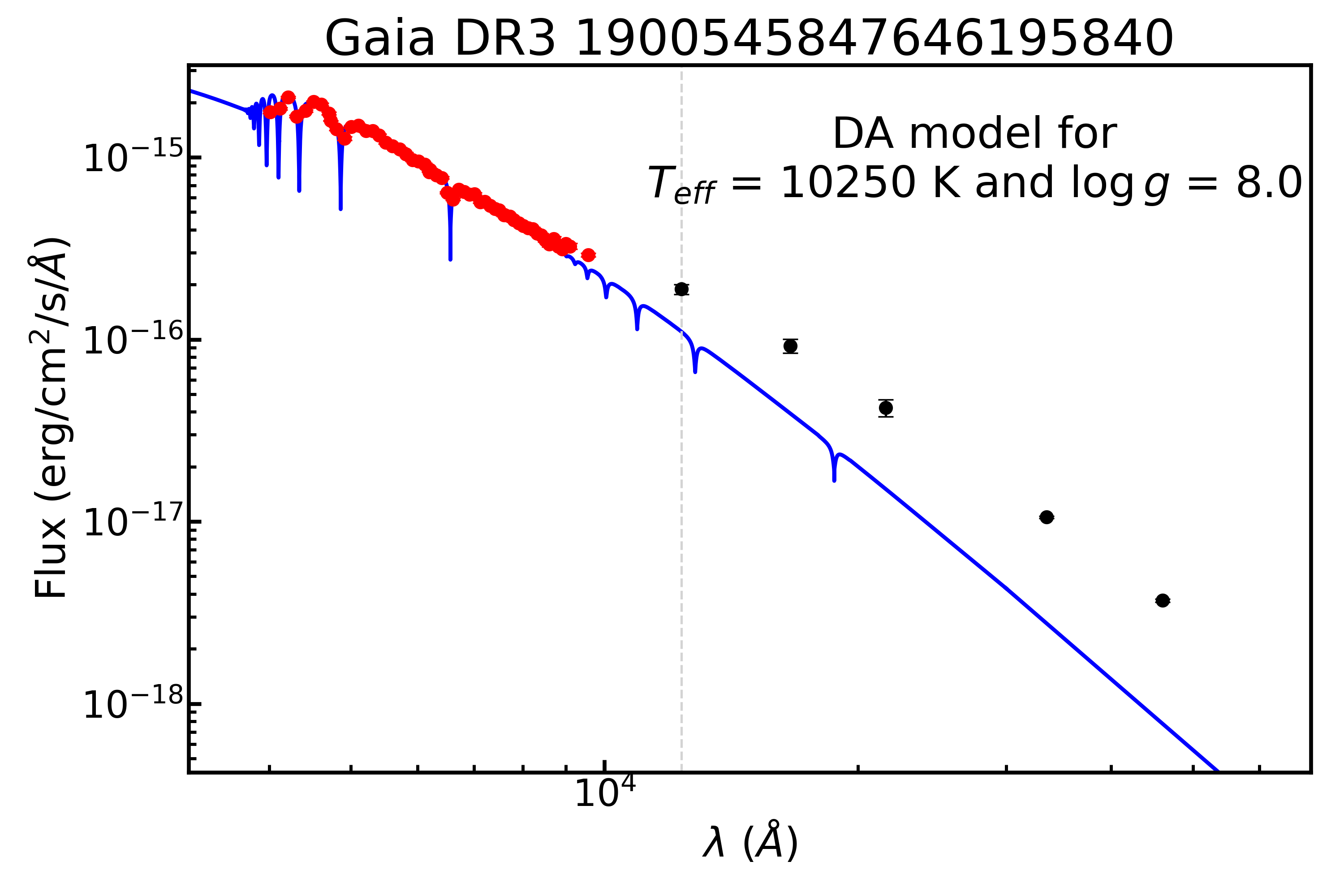}
    \includegraphics[width=0.35 \linewidth]{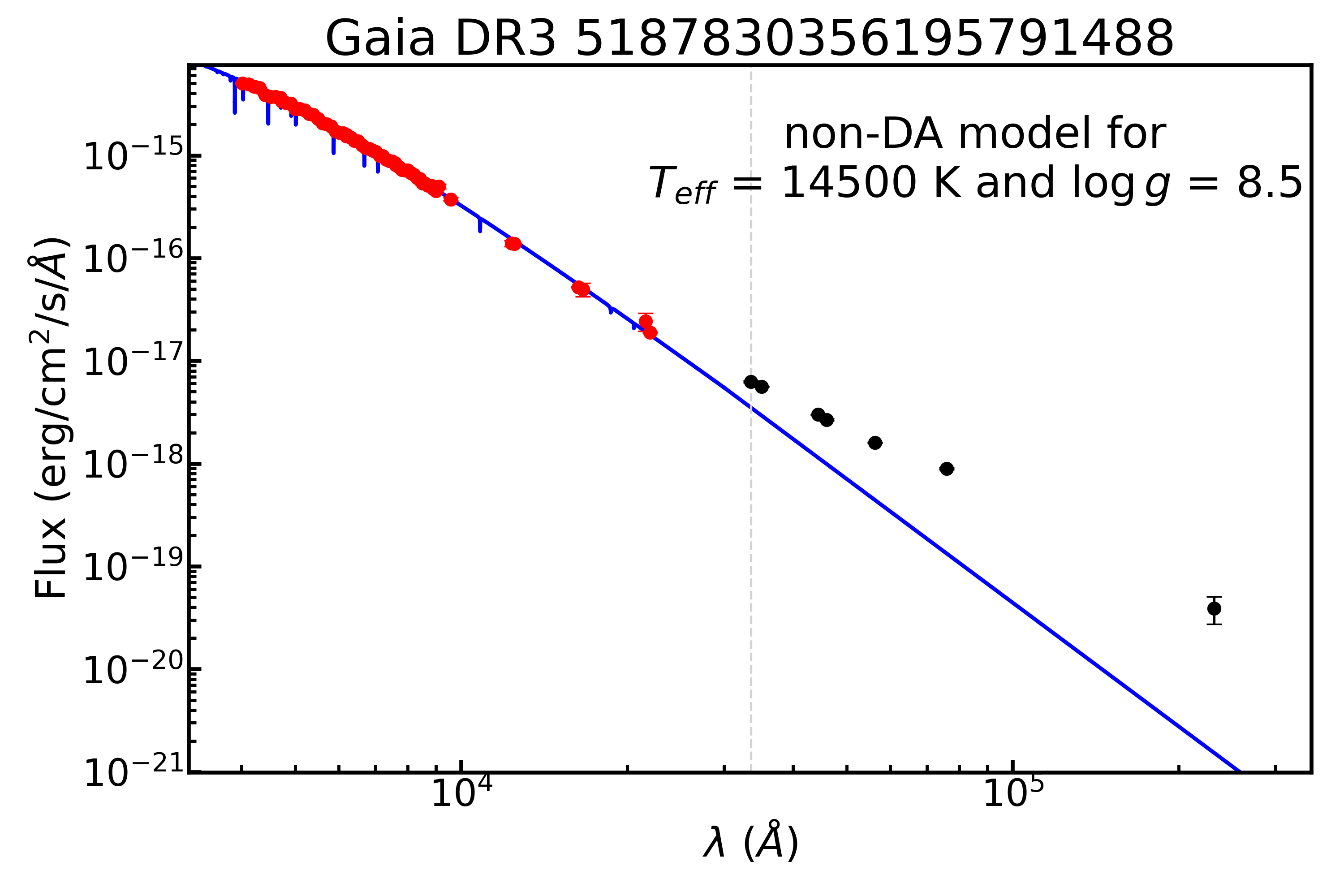}
    \caption{\small{SEDs of two white dwarfs (\texttt{Gaia DR3 1900545847646195840} and \texttt{Gaia DR3 5187830356195791488}) showing infrared excess emission. The left-hand panel shows an example of a DA model fit, and the right-hand panel an example of a non-DA model fit. Models are shown with blue lines, and the photometric data with black or red points depending on whether excess flux emission with respect to the model is detected or not, respectively.}}
    \label{fig:best_fit}
\end{figure*}

Figure \ref{fig:best_fit} shows an example of a good fit to a DA model (left-hand panel) and to a non-DA model (right-hand panel) with infrared excess emission. 

After this process, we selected only those SEDs with at least two photometric points showing excess emission, which made a total of 674 sources.

\subsection{Cleaning up of photometric contamination}
\label{contamin}

The final step in building the sample was a visual inspection of astronomical images to identify mismatched counterparts or contaminated photometry in the infrared catalogues for the 674 sources identified in the previous step. For this task, we used \texttt{Aladin}\footnote{\url{https://aladin.u-strasbg.fr/}} \citep{2000A&AS..143...33B}, one of the most widely used VO tools. It is an interactive sky atlas that allows the visualisation of astronomical images and the overlaying of catalogue data. 

\begin{figure*}[ht]
    \centering
    \includegraphics[width=0.375 \linewidth]{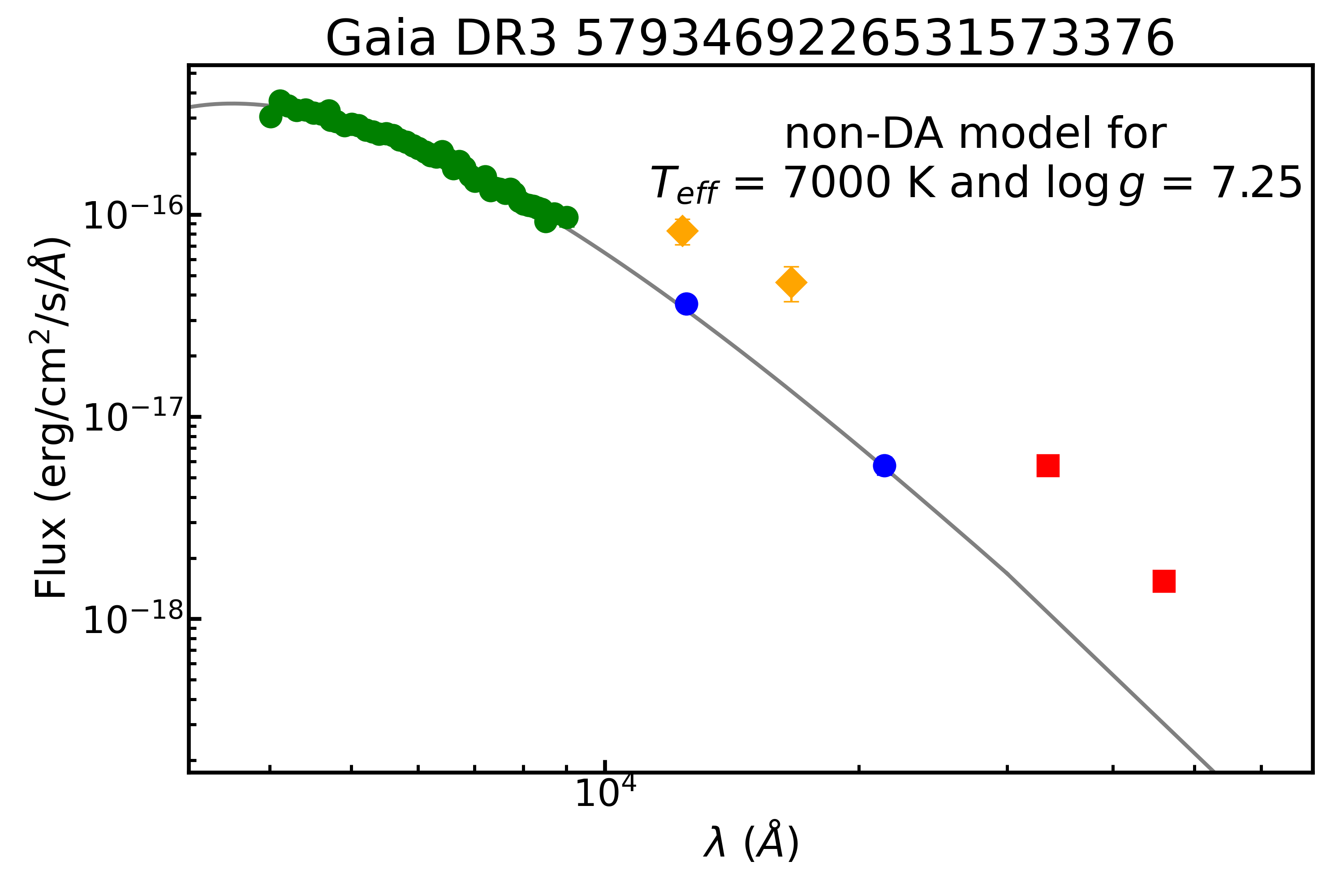}
    \includegraphics[width=0.3 \linewidth]{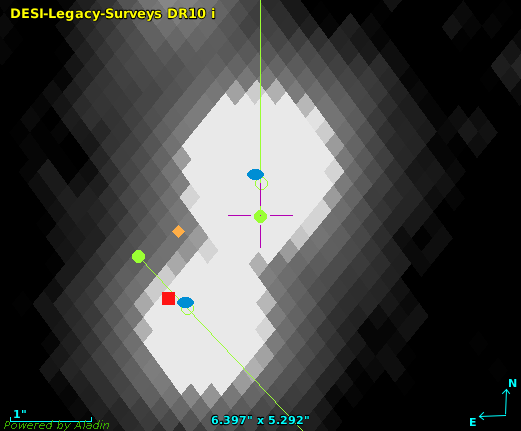}
    \includegraphics[width=0.3 \linewidth]{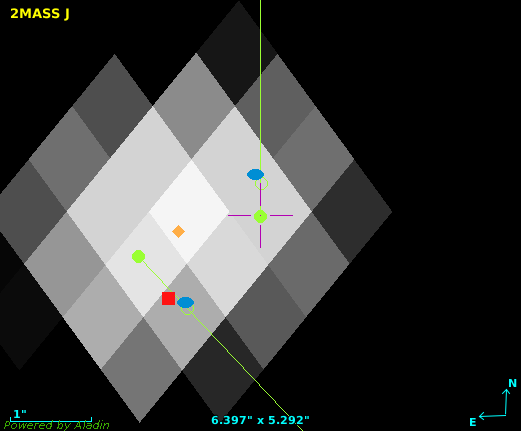} 
    \caption{Example of a white dwarf SED with contaminated photometry, \texttt{Gaia DR3 5793469226531573376}. \emph{Left panel:} SED with the photometric data shown with different colours and symbols: J-PAS with green points, 2MASS with orange diamonds, VISTA with blue ovals, and WISE with red squares. The grey line corresponds to the non-DA model that best fits the photometric data. \emph{Middle and right panels:} DESI (optical) and 2MASS (near-infrared) images, respectively, centred at the \G{} DR3 position of the white dwarf at epoch J2000 (magenta cross), and showing the white dwarf and a nearby field star at the southeast. 
    \G{} DR3 counterparts at epochs J2000 and J2015.4 are shown with green solid and empty circles, respectively. \G{} DR3 proper motion directions are shown with green lines. The line running from south to north is for the white dwarf, while the line running from northeast to southwest is for the nearby field star. The 2MASS counterpart is shown as an orange diamond, the WISE counterpart is shown as a red square, and the VISTA counterparts are shown as blue ovals.}
    \label{fig:cont}
\end{figure*}

Thus, we visualized images of different surveys in the optical (Dark Energy Spectroscopic Instrument --DESI-- Legacy Imaging Survey \citep{2019AJ....157..168D}, PanSTARSS \citep{2016arXiv161205560C}, Sloan Digital Sky Survey Data Release 9 (SDSS9, \citealt{Ahn+12}) and Digitized Sky Survey Data Release 2 (DSS2, \citealt{Laster+1996})) and in the infrared (2MASS, UKIDSS, VISTA, WISE and Spitzer). Additionally, for each white dwarf, we overplotted the \G{} DR3 counterpart together with the counterparts at the different infrared catalogues used to build the object SED. 
For illustrative purposes, in Figure \ref{fig:cont}, we show the DESI optical i-band and 2MASS infrared J-band images of \texttt{Gaia DR3 5793469226531573376}, which had a contaminated SED. The optical image clearly shows a field source near the white dwarf, which is not resolved in the 2MASS image. Thus, in contrast to \G{} and VISTA, the photometric measurements of 2MASS and WISE, which have worse resolution power than 2MASS, are the combined emission of both objects. After discarding the 2MASS and WISE data, the SED did not show any excess emission.

During the visual inspection of the available images at \texttt{Aladin}, which was essential in the identification of photometric contamination due to nearby objects, we labelled the SEDs in three categories: contaminated photometry, reliable excess emission, and tentative excess emission. In the case of identification of contaminated photometry, we dropped the corresponding photometry data from the SEDs and reanalysed them with \V{} to check if the SEDs continued showing excess emission or not. After this process, we rejected 218 sources whose corrected SEDs no longer showed excess emission (65) or had insufficient ($<$3) infrared photometry points (153).

\section{The infrared excess white dwarf sample}
\label{sec:Results}

The whole process described in the previous section ended up with a sample of 456 white dwarfs displaying infrared excess (hereafter `the infrared excess sample'). 
Among these, 292 show a reliable excess detection (hereafter `the reliable sample'), including 157 DA and 135 non-DA white dwarfs. The remaining 164 white dwarfs present a tentative excess (hereafter `the tentative sample'), with 117 classified as DAs and 47 as non-DAs.
The whole infrared excess sample is publicly accessible at the CDS VizieR catalogue service\footnote{\footnotesize \url{https://vizier.cds.unistra.fr/viz-bin/VizieR}} and at \emph{The SVO archive of White Dwarfs from Gaia}\footnote{\url{http://svocats.cab.inta-csic.es/wdw/}}, an online archive hosted at the Spanish Virtual Observatory (SVO) website.
Instructions on how to access the catalogue and a table with the list and the description of the parameters provided for each object are available in Appendix \ref{sec:appendixA}.

\begin{figure*}[ht]
    \centering
    \includegraphics[width=0.49 \linewidth]{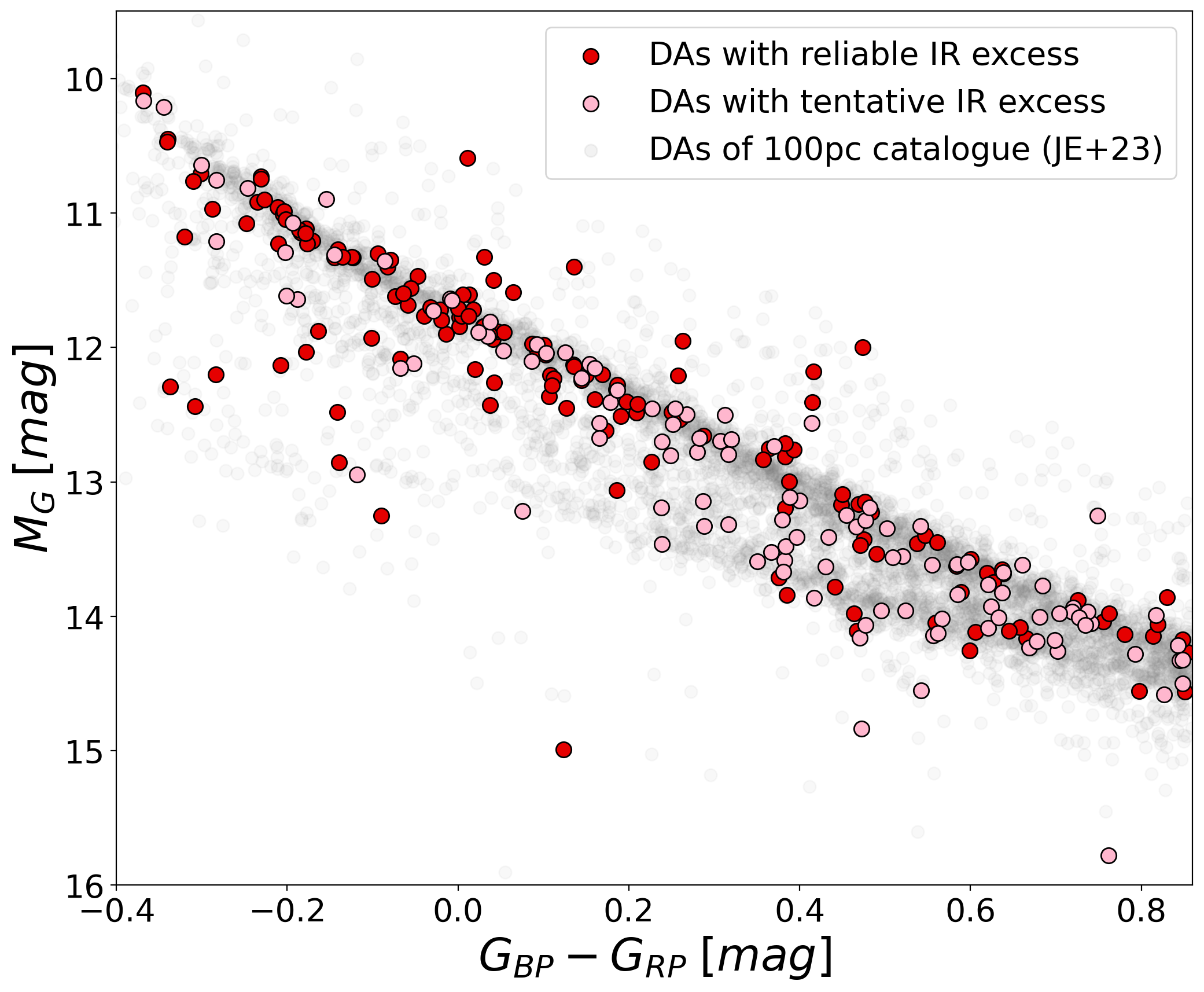} 
    \includegraphics[width=0.49 \linewidth]{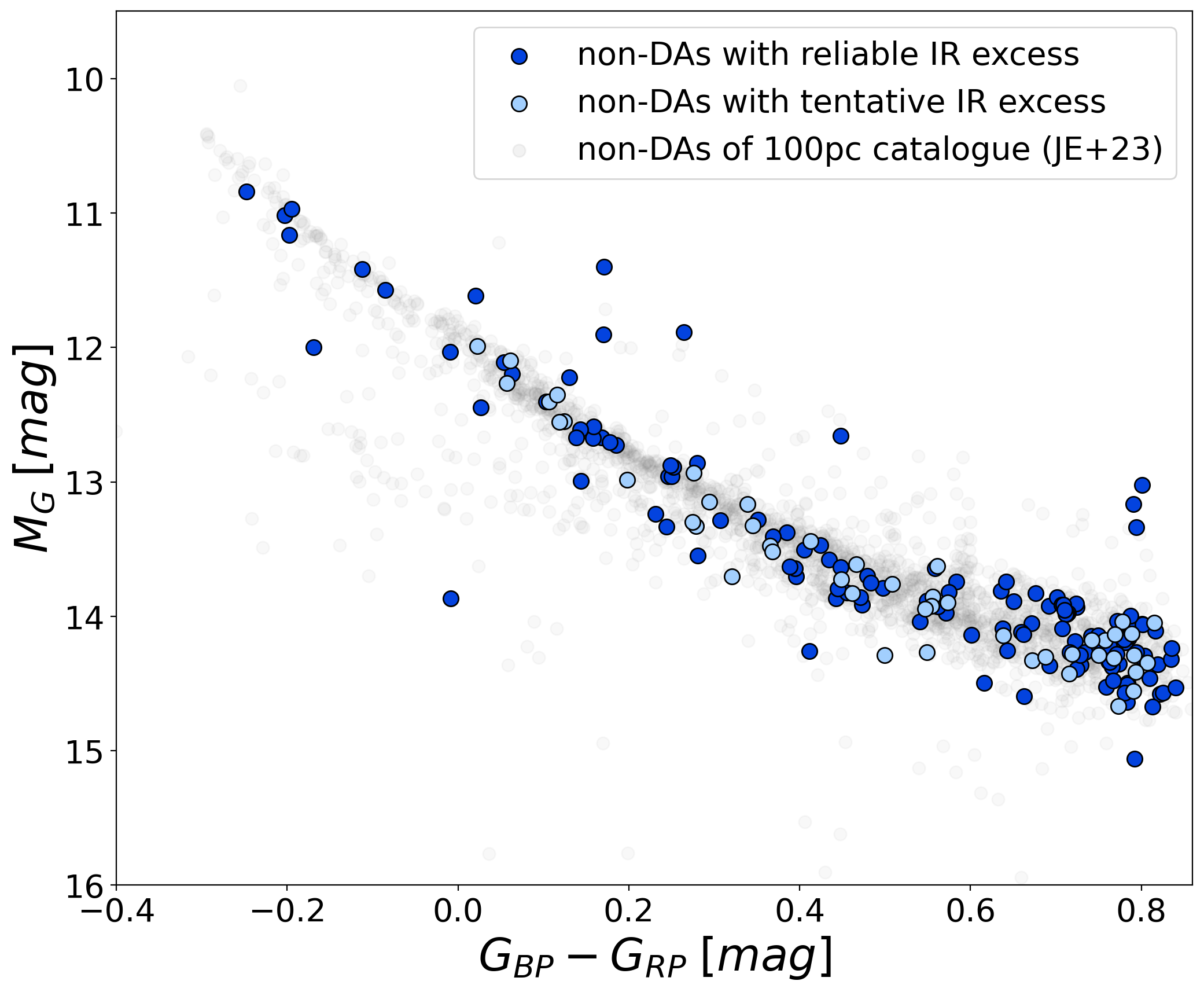} 
    \caption{\G{} Hertzsprung-Russell diagrams for the DAs (left panel) and non-DAs (right panel) with reliable and tentative infrared excess detections. For comparative purposes, we show the 100\,pc white dwarf DAs (left panel) and non-DAs (right panel) population from \citetalias{2023MNRAS.518.5106J}.}
    \label{fig:HRD}
\end{figure*}

Figure \ref{fig:HRD} shows the infrared excess sample in the \G{} 
Hertzsprung-Russell (HR) diagram. DAs are shown on the left panel and non-DAs on the right panel. No significant differences are visible between the distribution of the sources with reliable (dark colour) and tentative (light colour) infrared excess identification, regardless of spectral type. The distribution of DAs and non-DAs with infrared excess is also similar to the corresponding general distribution of DAs and non-DAs populations of white dwarfs within 100\,pc (grey circles). Notably is the concentration of non-DAs showing infrared excess at the reddest colours, which is not present for the DAs.

\begin{figure*}[ht]
    \sidecaption
    \centering
    \includegraphics[width=0.35 \linewidth]{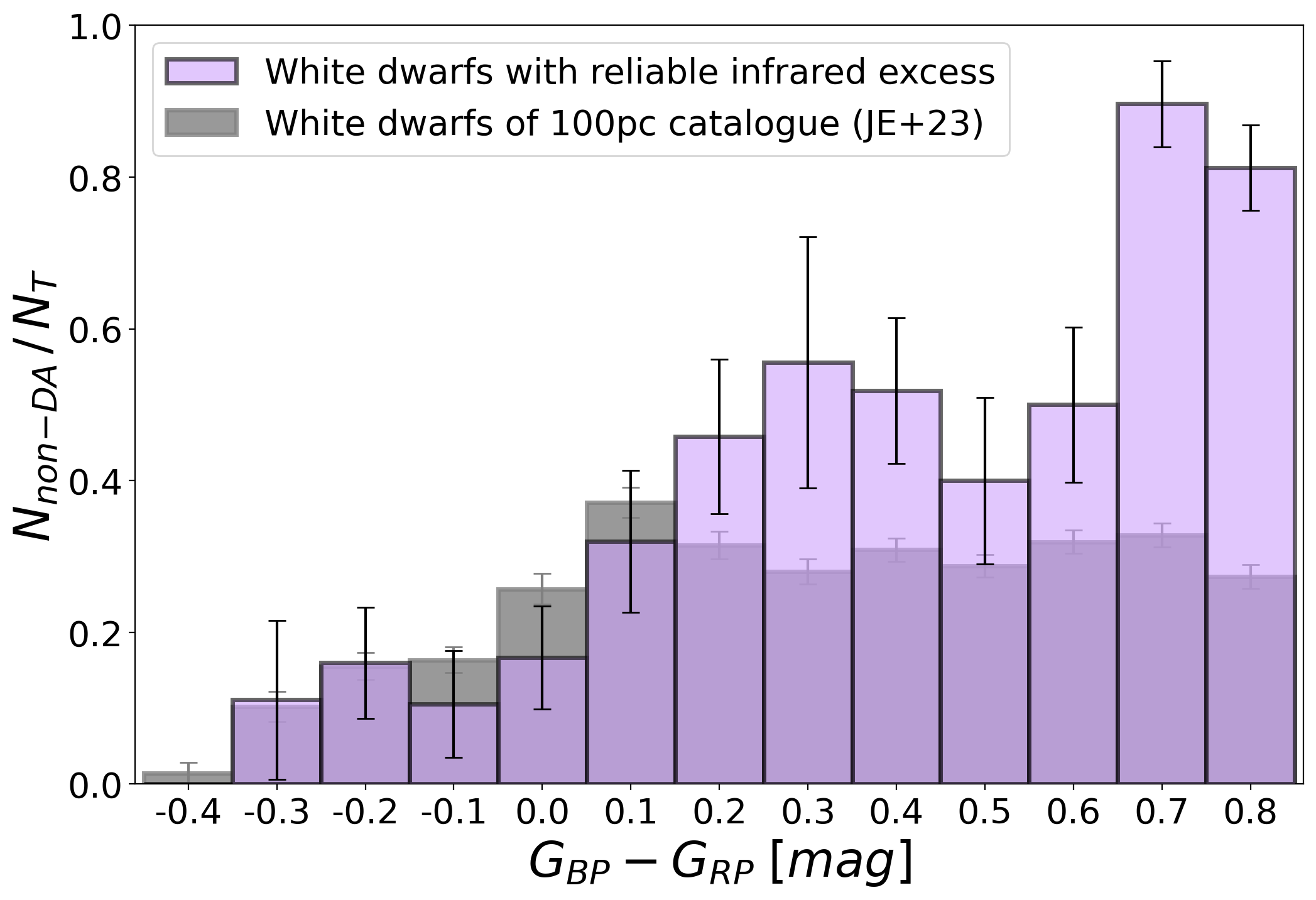} 
    \includegraphics[width=0.35 \linewidth]{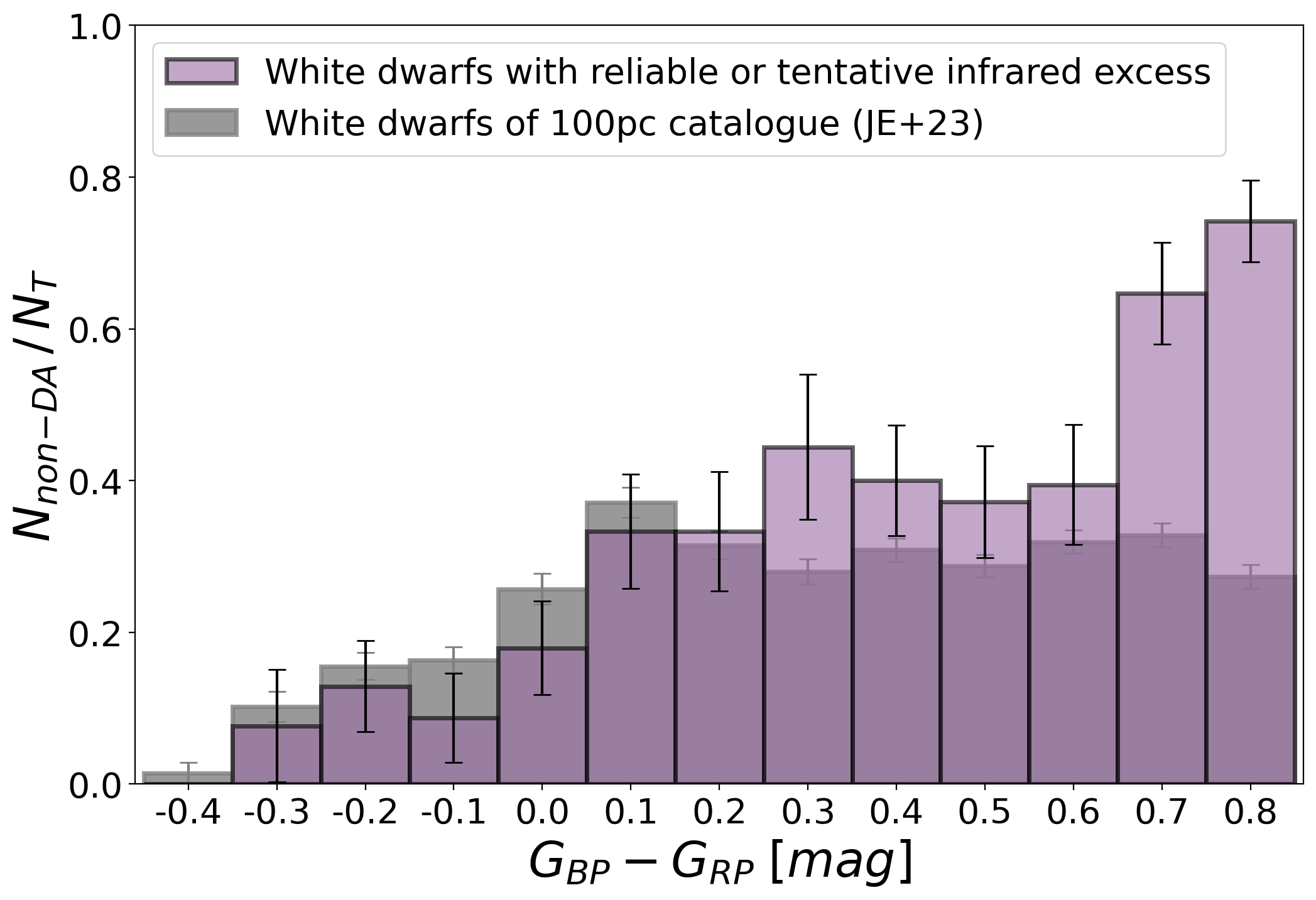} 
    \caption{Distribution of the fraction of non-DAs with infrared excess relative to the total number of white dwarfs with infrared excess as a function of the \G{} colour. The reliable sample is shown in the left panel, and the total infrared excess sample in the right panel. For comparative purposes, the same ratio for the general population of white dwarfs at 100\,pc is shown in grey.}
    \label{fig:colours}
\end{figure*}

Figure \ref{fig:colours} shows the ratio of non-DA white dwarfs with infrared excess relative to the total number of white dwarfs with infrared excess, $N_\mathrm{T}=N_\mathrm{DA}+N_\mathrm{non-DA}$, as a function of $G_\mathrm{BP}-G_\mathrm{RP}$ colour for the reliable sample (left panel) and the full infrared excess sample (right panel). Additionally, we show this ratio for the general population of white dwarfs at 100\,pc. The errors were calculated assuming a binomial distribution as $\sigma = \sqrt{\frac{f (1-f)}{N}}$, where $f$ is the fraction of positive cases and $N$ the total number of cases.

There was a clear trend of increasing ratio with increasing colour, less pronounced if we considered both the reliable and the tentative detections (right panel), but still very clear. This is consistent with the overdensity of non-DAs at redder colours in Figure \ref{fig:HRD}. Although this first analysis indicates an increasing ratio of infrared excess among non-DAs compared to the general population, the result should be interpreted with caution, especially for the reddest bins. This may reflect observational or methodological biases affecting the faint end of the sample. Non-DA atmospheric models, particularly for cool helium-rich atmospheres, are subject to larger uncertainties in the infrared regime \citep[e.g][]{Kowalski2014,Blouin} and may therefore be more prone to apparent infrared excess than DA models. This effect could reflect limitations in current atmospheric modelling rather than a genuine astrophysical trend, and a more robust interpretation will require accurate determinations of stellar temperatures, to be presented in a forthcoming publication.

\begin{figure}[ht]
    \centering
    \includegraphics[width=0.98
    \linewidth]{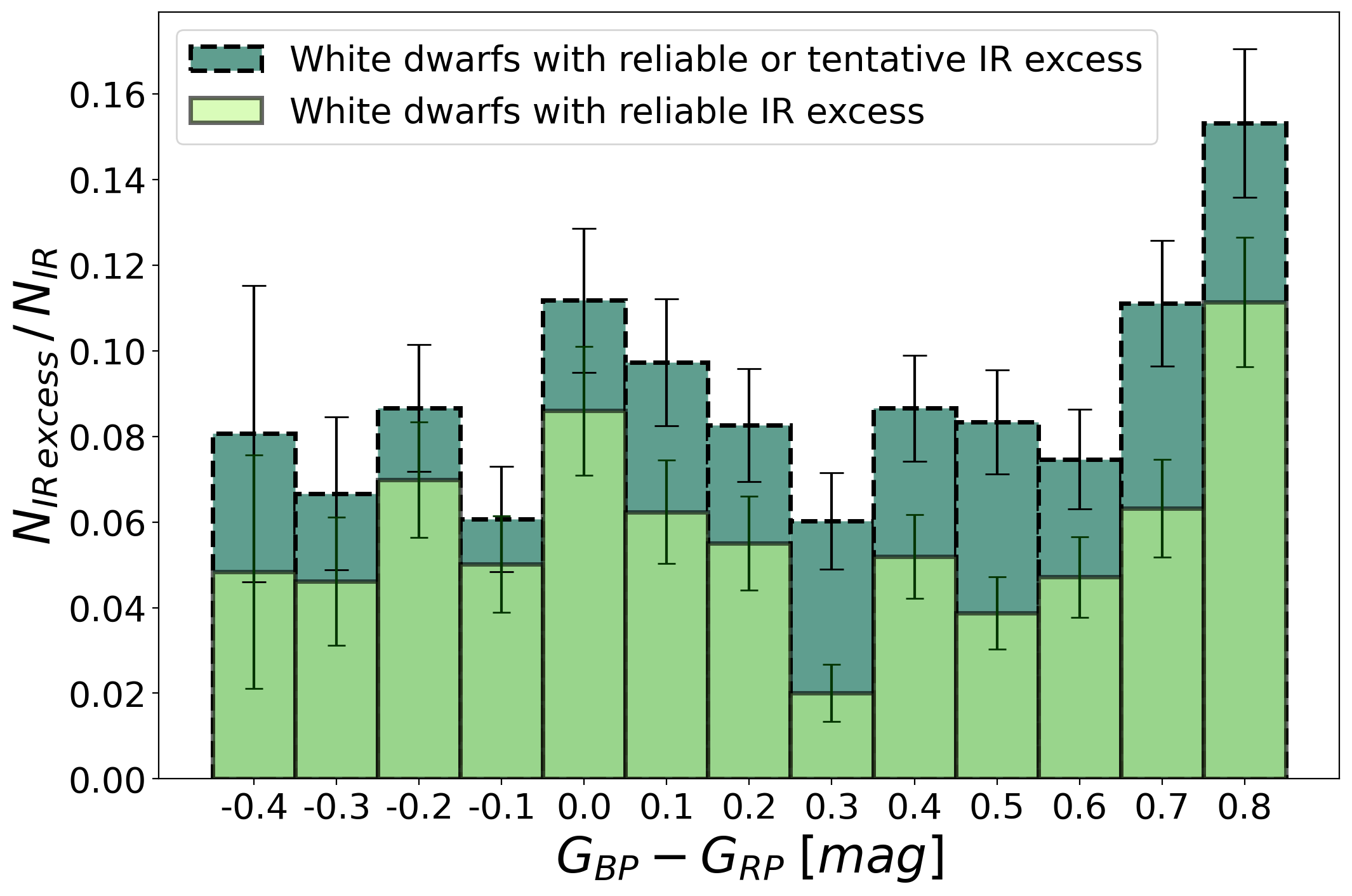}
    \caption{Distribution of the fraction of white dwarfs with infrared excess with respect to the number of white dwarfs with enough infrared photometry data in function of the \G{} colour.}
    \label{fig:rel_ir}
\end{figure}

Finally, Figure \ref{fig:rel_ir} shows the ratio of white dwarfs with infrared excess (both tentative plus reliable and only reliable) with respect to the total number of white dwarfs with enough infrared photometric data to be searched for infrared excess in this work, and after excluding the contaminated photometry, as a function of the $G_\mathrm{BP}-G_\mathrm{RP}$ colour. While the ratio remained relatively constant, a noticeable increase was observed at the redder end of the distribution.
This is in line with some studies that suggest that asteroid or planetesimal disruption may occur over long timescales, potentially affecting older white dwarfs \citep[e.g.][and references therein]{VerasHeng2020}, although this trend is not firmly established \citep[e.g.][]{Xu_20}. Moreover, cooler white dwarfs emit less flux, making even a small infrared excess easily detectable. However, this apparent increase in the redder bins of Fig.\,\ref{fig:rel_ir}  may largely reflect the high number of non-DA white dwarfs with identified infrared excess in those bins, which should be interpreted with caution. In Sect.\,\ref{sec:fract}, we provide a more detailed analysis of the fraction of white dwarfs exhibiting infrared excess. 

\section{Comparisons with previous works}
\label{sec:Comparisons}
    
In this section, we compared our catalogue with other works found in the literature. These works were divided into two groups: those which used photometry to identify white dwarf candidates with infrared excess and those for which the infrared excess was spectroscopically confirmed.

\subsection{Photometric infrared excess catalogues}    
\label{sec:Comparisons_phot}

\subsubsection{Candidates from UKIDSS DR8}

Two works by \cite{Steele+11} and \cite{Girven+11} used UKIDSS DR8 near-infrared photometry to search for substellar companions and debris disk candidates around white dwarfs. 
In the first work, the search was based on two spectroscopically identified white dwarf catalogues, while the second was based on a colour selected white dwarf candidate sample built from the SDSS catalogue. They identify white dwarfs with infrared excess emission by comparing the observed UKIDSS photometry with the expected white dwarf emission estimated from atmospheric models.

\cite{Steele+11} found 33 (Table 7 of their paper) and \cite{Girven+11} found 109 (Table 5 and Table 6 of their paper) candidates for infrared excess. Among them, there are only two at less than 100\,pc, actually both sources identified in both works. We confirmed the infrared excess emission of both sources, as they are in our reliable sample.

\subsubsection{Candidates from Spitzer and AllWISE}
\label{sec:Dennihy+20}

The study by \cite{Dennihyelal2020_WISE} examined a sample of 22 white dwarfs previously reported to exhibit infrared excesses based on the AllWISE survey. Using follow-up observations with Spitzer, they confirmed that eight of them had AllWISE contaminated photometry by an unresolved field source.

Seven of the objects in their initial sample lie within 100 pc. Of these, three were classified as genuine infrared excesses, while the remaining four were identified as contaminated. All seven objects also appear in \citetalias{2023MNRAS.518.5106J} and are included in our sample. Our classifications are fully consistent with theirs: the three objects with confirmed excesses are part of our reliable sample, whilst the other four show no indication of an excess in our analysis. One important difference between the two studies is the choice of WISE catalogues. We used CatWISE, whereas they used AllWISE. As explained in Sect.\,\ref{sec:fract}, CatWISE is a much better choice for white dwarfs.

\subsubsection{Candidates from the WIRED survey}

The WISE InfraRed Excesses around Degenerated (WIRED) survey was developed to detect infrared excess around white dwarfs from WISE photometry. In a series of papers \citep{Debes+11a,Debes+11b,Hoard+13,Dennihy+17}, the authors analysed samples of white dwarfs from the SDSS DR7 WD catalogue \citep{Kleinman10}, the McCook \& Sion white dwarf catalogue \citep{McCook&Sion+99}, and Edinburgh-Cape Blue Object Survey \citep{Stobie+97}, and used Preliminary, All-Sky, 3-Band Cryo or AllWISE Data Release of WISE to identify infrared excess candidates.

WIRED authors identified a total of 1296 white dwarfs with infrared excesses. Of them, 111 were within 100\,pc, 90 were in \citetalias{2023MNRAS.518.5106J}, but six without classification and nine without enough infrared photometry to be included in our analysed sample. Of the remaining 75 sources in common, 33 are in the reliable sample and three in the tentative sample, we identify another three with contaminated photometry. The remaining 36 showed no excess. 

The large percentage of sources for which we did not find infrared excess is probably due to misidentification of their infrared counterpart and to contaminated photometry, especially in WISE. They did not correct the white dwarf position for proper motion and used a larger search radius, which resulted in wrong identifications. In addition, they relied on earlier versions of the WISE catalogues that contained poorer photometry with many more blended sources, as explained above.

\subsubsection{Candidates from \G{} DR2 and the VO} 

\cite{RM+19} used the VO to search for infrared excess white dwarfs within a 100\,pc white dwarf sample built from the \G{} DR2 catalogue \citep{JE+18}, paving the way to the present work. Since no \G{} spectra were published in the DR2, they built the optical part of the object's SED using photometric catalogues available at the VO, and similarly to this work, used \V{} to fit the SEDs with DA models and identify the excesses.

\cite{RM+19} identified 77 white dwarfs with infrared excess (Table 3 and Table 4 of that paper). Of them, 58 sources were also identified as having infrared excess in this work, 55 in our reliable sample and three in the tentative sample. Of the remaining, ten showed no excess, eight were identified in the images inspection process as having contaminated photometry, and one did not have enough infrared photometry for our analysis.

Additionally, \cite{RM+19} presented a sample of 33 white dwarfs with tentative infrared excess in their Table 5. Of them, we identified three with reliable and two with tentative infrared excess. The remaining 27 showed either no excess (two), contaminated photometry (two), not enough infrared photometric points (20), or were not in our initial sample (four).

These discrepancies are likely due to several improvements we introduced in this work: 
\begin{itemize}
    \item Our studied sample comes from the 100\,pc white dwarf sample by \citetalias{2023MNRAS.518.5106J}, which was built from \G{} DR3 instead of DR2 thus increasing the number of white dwarfs, especially at the redder colours. Additionally, white dwarfs are now split between DAs and non-DAs using their \G{} spectra, which were not available before.
    
    \item The optical part of the SEDs has greater and better coverage with the synthetic photometry of J-PAS, derived from \G{} DR3.
    
    \item Nowadays, infrared catalogues such as UKIDSS or VISTA have larger sky coverage, providing infrared data for a greater number of white dwarfs than before. Additionally, we used \G{} DR3 proper motion to translate the source coordinates into the catalogue epoch, which permitted us to use a smaller search radius of 3\arcsec. Thus, we increased the number of identified infrared counterparts while reducing the number of miss-identifications.

    \item The mid-infrared photometry data from CatWISE used in this work superseded the AllWISE photometry data used in \citet{RM+19}, which greatly improved the quality of the mid-infrared data (see Sect.\,\ref{sec:fract}). 
    
    \item We used DAs and non-DAs white dwarf atmospheric models to estimate with \V{} the expected emission of the source, which allowed a more reliable infrared excess identification. In fact, of the 180 non-DA white dwarfs with infrared excess identified in this work, less than half (79) would have been identified if the SEDs would have been fitted with DA models. Thus, using non-DAs models clearly increased the number of infrared excess identifications. 

    \item We extended the $G_\mathrm{BP}-G_\mathrm{RP}$ colour range from 0.8 to 0.86, where the fraction of infrared excess white dwarfs increases considerably (see Fig.\,\ref{fig:rel_ir}).

\end{itemize}

All of these allowed us to detect not only a greater number and more reliable infrared excesses, but also rule out previous spurious identifications.

\subsubsection{Candidates from \emph{Gaia} DR2 and unWISE}
\label{sec:Xu_Lai}

\cite{Xu_20} presented a sample of 188 infrared excess candidates around bright white dwarfs obtained from \G{} DR2 and unWISE catalogues.  
In parallel, \cite{Lai_21} used Spitzer to search for infrared excess around 174 white dwarfs (Table B1 and Table B2 of that paper), 56 of them in common with the \cite{Xu_20} sample. They confirmed 62 white dwarfs with infrared excess, 44 of them in common with \cite{Xu_20} sample, but the remaining 18 objects were previously rejected by \cite{Xu_20}.

Of the 174 initial candidates from \cite{Lai_21}, 79 are closer than 100\,pc. Of them, 69 were included in \citetalias{2023MNRAS.518.5106J}, but one was not spectrally classified, and eight were discarded because they did not have enough infrared photometry. Thus, there were 60 common objects between our initial sample and the initial sample of \cite{Lai_21}. We agreed that 16 of them had reliable infrared excess and 28 showed no excess; the remaining 16 were in our infrared sample (nine reliable and seven tentative), but were classified with no excess by \cite{Lai_21}. 
This may be due to several factors. First, they applied a more restrictive condition to define a flux excess, the difference between the observed and model flux had to differ at less 3.27 or 4.80 times the error module for Spitzer IRAC Ch1 (3.6$\mu$m) and Ch2 (4.5$\mu$) bands, respectively. Second, their optical part may not be well characterised due to the limited data of their SEDs and the use of only DA atmospheric models.

The \cite{Xu_20} sample had 188 objects, 72 at less than 100\,pc; 67 were classified in \citetalias{2023MNRAS.518.5106J}, but six of them lacked sufficient infrared photometry. 
Thus, we had 61 in common; we classified 42 with reliable excess, including the ten objects mentioned above, and two with tentative excess. However, we found that one object presented contaminated photometry, and another 16 showed no infrared excess.
Moreover, of the sources that \cite{Xu_20} rejected as contaminated photometry, we identified 14 of them with either reliable (six) or tentative (eight) infrared excesses. Even more, we were able to identify another 48 sources with infrared excesses, 40 reliable and eight tentative, which did not fulfil the criteria imposed by \cite{Xu_20} to be included in their infrared excess candidate sample. 

The different results obtained are mainly due to the use of improved data, with a better characterisation of the optical part of the white dwarf SED from the \G{} DR3 spectra, more reliable infrared photometric data from CatWISE instead of unWISE catalogue, and a more exhaustive identification of the contaminated photometry.

\subsubsection{Candidates from LAMOST and WISE}

In \cite{Wang23}, a sample of \WDs{} with spectroscopic data in LAMOST DR5 and with distances in \G{} DR2 was analysed for infrared excess in WISE (AllWISE and CatWISE). They identified 52 infrared excess candidates (Table 1, Table 2 and Table 3 of their paper). Of these, only two white dwarfs are within 100\,pc, and both were included in our analysed sample. One showed a reliable infrared excess, and the other showed no excess. The difference was due to our better SED coverage in the optical wavelength range.

\subsubsection{Candidates from \G{} DR3 and CatWISE}

Two recent works, \cite{Madurga_Favieres_24} and \cite{Bravo+25}, based their search for infrared excess in \G{} DR3 and CatWISE, as we also did. Therefore, one would expect similar results, although this was not the case.

\cite{Madurga_Favieres_24} presented a candidate sample of 554 white dwarfs with infrared excesses identified using \G{} EDR3 and CatWISE, 64 of which were within 100\,pc. Of these, 62 were in the \citetalias{2023MNRAS.518.5106J} catalogue; however, two without spectral classification, and eight did not have enough infrared photometry. The remaining 52 were studied in this work, with the following result: 40 showed reliable and four tentative infrared excess, in two cases the photometry was contaminated, and the remaining six showed no excess. Additionally, many sources that we identified as showing infrared excess (144 reliable and 74 tentative) were discarded by \cite{Madurga_Favieres_24}. These discrepancies arise because very restrictive conditions were imposed: the difference between the observed and model magnitudes, taking into account the errors, had to be greater than eight or 30 times, depending on the apparent G magnitude of the object.

The second work mentioned above, \citet{Bravo+25}, presented a candidate sample of 77 \WDs{} with infrared excess (Table 1 and Table 2 of their paper), identified as part of a search for ultracool companions to \WDs{}. 
In this case, the search was limited to a distance of 100\,pc from the Sun.

Of the 77 identified excesses, 27 were already identified by \cite{Madurga_Favieres_24}. 
Of the remaining 50 objects, only 28 were in the \citetalias{2023MNRAS.518.5106J} catalogue, but four with no spectral classification and five did not have sufficient infrared photometry. Thus, we had 19 objects in common, 15 in the reliable sample and one in the tentative sample, one had contaminated photometry, and two showed no excess.

\subsection{Spectroscopically confirmed infrared excess}
\label{sec:Comparisons_spec}

We reviewed the literature to identify whether any of our infrared excess white dwarfs had been studied in detail, in order to determine if our results are consistent with previous spectroscopic works. In Table \ref{tab:spec_confirmed} we present a bibliographical compilation of \WDs{} with spectroscopically confirmed excess emission in the solar neighbourhood (< 100\,pc). The table contains the \G{} DR3 identifier, the excess type classification in this work, the reference in which the excess was confirmed spectroscopically, the system type spectroscopically defined, and the telescope and instrument used in each case.

Recently, \cite{Limbach+24} and \cite{Farihietal2025} confirmed with the James Webb Space Telescope (JWST) the existence of some white dwarfs with infrared excess for which we did not detect any. However, in these cases listed at the bottom of Table\,\ref{tab:spec_confirmed}, the infrared excesses started at longer wavelengths than the available photometry used in this work.

In conclusion, we were able to identify the infrared excess in the photometry of all the white dwarfs with infrared excess confirmed spectroscopically, as long as there is available photometry at the wavelengths at which the excess exists.

\section{The fraction of infrared excess white dwarfs}
\label{sec:fract}

An important aspect to consider is the overall fraction of white dwarfs that exhibit infrared excess. After cleaning the contaminated photometry (see Sect.\ref{contamin}), we searched for infrared excess in a total sample of 4931 white dwarf SEDs with at least three good infrared photometric points ($\lambda>12\,000\,$\AA; see Sect\,\ref{sec:meth}). Of them, we identified 292 reliable and 164 tentative infrared excesses. Thus, the estimated overall fraction of infrared excess white dwarfs was 9.2$\pm$0.4\%, or 5.9$\pm$0.3\% when considering only reliable excesses.

As shown in Fig.\,\ref{fig:rel_ir}, the largest rates of infrared excess white dwarfs concentrate in the reddest colours. However, even if we excluded the reddest objects ($G_\mathrm{BP}-G_\mathrm{RP}$>0.66), the percentages of tentative and reliable infrared excess and of only reliable excess sources are 8.3$\pm$0.4\% and 5.3$\pm$0.4\%, respectively, still higher than most of the values found in the literature ($\lessapprox$4\%) \citep[e.g.][]{RM+19, Wilsonetal2019, Rogersetal2020, Wang23}. A consideration arises from the fact that all but \cite{RM+19} samples were magnitude-limited, which directly affects their completeness, particularly at lower temperatures. It was worth noting that \cite{Xu_20} calculated a ratio of 6.6$\pm$0.5\%, which increased up to 8.4$\pm$0.6\% when taking into account completeness issues. Similarly, \cite{Madurga_Favieres_24} obtained a fraction of $\approx$ 6\%. These values are closer to those obtained in our study, based on the nearly-complete volume-limited 100\,pc sample of \citetalias{2023MNRAS.518.5106J}.

One of the main obstacles when estimating the percentage of infrared excess white dwarfs is the assessment of the contamination by field sources, especially when the identification of the excess emission was based on WISE data. \cite{Dennihyelal2020_WISE} used Spitzer imaging to confirm the infrared excess of a sample of AllWISE selected sources, finding $\approx$36\% of contamination on a sample of 22 candidates (see Sect.\,\ref{sec:Dennihy+20}). \cite{Lai_21} used Spitzer photometry to confirm the infrared excess of a sample of 176 white dwarfs selected from unWISE photometry. 
In this case, the rate of contamination decreased to $\approx$20\%. These better results could be explained by the improvement of the used WISE data. UnWISE is based on significantly deeper imaging that combines five years of observations instead of just thirteen months as AllWISE. Furthermore, unWISE also improves the modelling of crowded regions \citep{Schlaflyetal2019}.

We used CatWISE for our study, which improves upon unWISE by incorporating proper motions, by being based on eight years of observations, and by being specifically optimised for identifying and resolving faint sources. All this makes CatWISE much more reliable than any other WISE catalogue, especially for faint and high proper motion sources like white dwarfs. Consequently, we expect lower contamination than in the previous works.

We based the infrared excess detection on all available infrared photometry in the VO  (see Sect.\,\ref{sec:IRphot}). In fact, for $\approx$25\% of our sample, the infrared excess was detected in other photometry bands besides WISE ones. Assuming a similar contamination rate than with unWISE, and taking into account that $\approx$75\% of our sample, the excess is detected only on WISE data, our sample would be affected by $\approx$15\% of contamination. However, as we have explained above, CatWISE photometry is expected to be much less affected by background sources than the previously used WISE catalogues, so the expected contamination rate should be also much lower. Even more, we performed a very careful identification of contaminated photometry (Sect.\,\ref{contamin}), which resulted in the rejection of 218 candidates.

This is supported by the result of the comparison with the two above mentioned works in Sect.\,\ref{sec:Comparisons}. There is a full agreement in the classification with \cite{Dennihyelal2020_WISE} for the seven sources in common (see Sect.\,\ref{sec:Dennihy+20}). 
Furthermore, we agree with the classification of \cite{Lai_21} for 44 of the 60 common objects (see Sect.\,\ref{sec:Xu_Lai}). 
However, they applied a much restricted criterion to classify a source as with infrared excess emission, which could have resulted in missing true infrared excess sources. If they would have applied the same criteria as we did, they would have included another three common sources in their infrared excess sample. 
Taking these 3 sources and combining both works, the contamination would be $\approx$ 19\% for our infrared excess detections based on WISE data. However, as explained above, only 75\% of the detections were only based on WISE data, so we can set a maximum limit of 15\% for the overall contamination rate for our whole sample of infrared excess white dwarfs.
Even more, all the known white dwarfs with spectroscopic confirmation of infrared excess emission are in our infrared excess sample.

For all the reasons discussed above, we conclude that the contamination rate in our infrared excess white dwarf sample is low, with an upper limit of 15\%.

\section{Conclusions}
\label{sec:Conclusions}

We searched for sources with infrared excess emission in the 100\,pc sample of spectroscopically classified white dwarfs in \citetalias{2023MNRAS.518.5106J}. For this, we used VOSA to analyse their SEDs built from public spectroscopic and photometric archives. The optical part of the SEDs was populated with the synthetic J-PAS photometry obtained from \G{} DR3 low-resolution spectra, which provided a high coverage with 56 photometric points in the wavelength range between 4000 to 10\,500\,\AA. The infrared part of the SEDs was built from photometric catalogues available at the VO. In order to maximise the number of infrared counterparts while minimising the number of misidentifications, we used \G{} proper motion to translate the white dwarf coordinates at the epoch of each infrared survey and a small search radius of 3\arcsec.
We imposed a minimum number of seven photometric points, four in the optical and three in the infrared, in an SED to be further analysed with VOSA, although $\sim$70\% of the SEDs contained 59 or more photometric points. This restriction resulted in a sample of 5084 SEDs being searched for infrared excess emission.

To identify the infrared excess, we used VOSA to fit the SEDs to white dwarf atmospheric models, using DA or non-DA models accordingly to the spectral classification of each source provided by \citetalias{2023MNRAS.518.5106J}. During the fitting process, VOSA automatically identified any photometric point affected by excess emission by comparing the observed brightness with the white dwarf brightness predicted from the best fitted model. In order to ensure the reliability of the infrared excess emission found, we imposed firstly a minimum of two photometric points flagged as such by VOSA, and secondly we performed an exhaustive check of possible photometry contamination by inspecting both optical and infrared images available at the VO. About one third of the infrared excesses were ruled out because of contaminated photometry.

As a result of this process, we identified 456 white dwarfs showing infrared excess emission, 292 reliable and 164 tentative. Among these, about 75\% are new identifications. A catalogue containing these sources is accessible online at the CDS VizieR catalogue service\footnote{\footnotesize \url{https://vizier.cds.unistra.fr/viz-bin/VizieR}} and at \emph{The SVO archive of White Dwarfs from Gaia}\footnote{\url{http://svocats.cab.inta-csic.es/wdw/}}.

We found infrared excess emission in 5.9$\pm$0.3\% of the white dwarf SEDs analysed with VOSA when only the reliable sample is taken into account, a value that increased to $9.2\pm$0.4\% when the complete infrared excess sample was considered. This fraction is higher than that typically found in the literature. The use of highly populated SEDs, with 56 photometric points in the optical and the latest version of the infrared catalogues, together with the use of different grids of atmospheric models for DAs and non-DAs sources, would explain the higher success of our methodology. 
Based on common sources among our study and other studies in the literature, we estimated an upper limit of 15\% on the contamination rate, although we actually expect a substantially smaller contamination due to the use of the latest catalogue of WISE survey, CatWISE, in combination with other infrared catalogues. 

A first analysis of the data reveals that the fraction of non-DAs with infrared excess relative to the total number of white dwarfs with infrared excess increases with the \G{} $G_\mathrm{BP} - G_\mathrm{RP}$ colour (Fig. \ref{fig:colours}), in contrast with the general population for which the fraction remains almost uniform for $G_\mathrm{BP} - G_\mathrm{RP}$\,>\,0.2\,mag. However, these results should be taken with caution given that non-DA atmospheric models are subject to larger uncertainties in the infrared regime at low temperatures.

After comparing our results with previous photometric studies (Sect.~\ref{sec:Comparisons_phot}), we found that 99 of the sources in our infrared-excess sample were already identified. Additionally, three objects in our list were previously reported in other photometric works not discussed above \citep{Becklin&Zuckerman88, Farihi2009, Farihietal2010, Farihietal2012}. We also identified 69 objects showing no excess and 12 with contaminated photometry that were incorrectly classified as exhibiting infrared excess in at least one previous study. Furthermore, 26 of our objects have been confirmed spectroscopically (Sect.~\ref{sec:Comparisons_spec}).
Of them, 21 were in our infrared-excess sample, and the other five were not included because the excess emission becomes apparent only at wavelengths longer than the photometric coverage of our data.

The sample of nearby (<\,100\,pc) white dwarfs with infrared excess presented here is the most complete and reliable available to date. Although spectroscopic confirmation is still required, our methodology—combined with careful treatment of contaminated photometry—makes this sample well suited for follow-up observations. A detailed characterisation of these systems will be presented in a forthcoming paper, offering new opportunities to investigate circumstellar disks, substellar companions, and the composition of accreted planetary material.

\section*{Data availability}
The \G{} white dwarfs with infrared excess catalogue is available at the CDS via anonymous ftp, under the identifier \url{https://cdsarc.cds.unistra.fr/viz-bin/cat/J/A+A/} adding volume/first\_page.
It can also be queried via an online service available at the Spanish Virtual Observatory (SVO) website (see Appendix \ref{sec:appendixA}). Table \ref{tab:Cat} contains information about the catalogue columns. Note that the SED photometry and graphics are only available on the SVO website. 

\begin{acknowledgements}
    
    We thank the anonymous referee for their valuable comments.
    RMO is funded by INTA through grant PRE-OBSERVATORIO.
    We acknowledge support from MINECO under the PID2023-148661NB-I00 grant and by the AGAUR/Generalitat de Catalunya grant SGR-386/2021. 
    This work has made use of data from the European Space Agency (ESA) mission {\it Gaia} \url{https://www.cosmos.esa.int/gaia}), processed by the {\it Gaia} Data Processing and Analysis Consortium (DPAC, \url{https://www.cosmos.esa.int/web/gaia/dpac/consortium}). Funding for the DPAC has been provided by national institutions, in particular the institutions participating in the {\it Gaia} Multilateral Agreement.
    This work has made use of the Python package GaiaXPy, developed and maintained by members of the {\it Gaia} Data Processing and Analysis Consortium (DPAC), and in particular, Coordination Unit 5 (CU5), and the Data Processing Centre located at the Institute of Astronomy, Cambridge, UK (DPCI).
    This publication makes use of the \texttt{SVO Filter Profile Service "Carlos Rodrigo"} and \V{}, developed under the Spanish Virtual Observatory (\url{https://svo.cab.inta-csic.es}) project funded by MCIN/AEI/10.13039/501100011033/ through grant PID2023-146210NB-I00. \V{} has been partially updated by using funding from the European Union's Horizon 2020 Research and Innovation Programme, under Grant Agreement nº 776403 (EXOPLANETS-A).
    This research has made use of "Aladin sky atlas" developed at CDS, Strasbourg Observatory, France. 
     We extensively made used of Topcat \citep{Taylor05}.
    Based on data from \emph{the SVO archive of White Dwarfs from \G{}} at CAB (CSIC-INTA).
     This publication makes use of data products from the Two Micron All Sky Survey, which is a joint project of the University of Massachusetts and the Infrared Processing and Analysis Center/California Institute of Technology, funded by the National Aeronautics and Space Administration and the National Science Foundation.
     This publication makes use of data products from UKIDSS and VISTA. 
     The UKIDSS project is defined in \cite{Lawrenceetal07}. UKIDSS uses the UKIRT Wide Field Camera (WFCAM; \cite{Casalietal07}) and a photometric system described in \cite{10.1111/j.1365-2966.2005.09969.x}. The pipeline processing and science archive are described in \cite{Irwin08} and \cite{Hamblyetal08}. 
     The VISTA Data Flow System pipeline processing and science archive are described in \cite{Irwinetal04}, \cite{Hamblyetal08} and \cite{refId0}. 
     This publication makes use of data products from the Wide-field Infrared Survey Explorer, which is a joint project of the University of California, Los Angeles, and the Jet Propulsion Laboratory/California Institute of Technology, funded by the National Aeronautics and Space Administration; and the Near-Earth Object Wide-field Infrared Survey Explorer (NEOWISE), which is a joint project of the Jet Propulsion Laboratory/California Institute of Technology and the University of Arizona.
     This work is based in part on observations made with the Spitzer Space Telescope, which was operated by the Jet Propulsion Laboratory, California Institute of Technology under a contract with NASA.

\end{acknowledgements}

\bibliographystyle{aa}
\bibliography{bib.bib} 

\begin{appendix}
\onecolumn
\section{Online catalogue service}
\label{sec:appendixA}

In order to help the astronomical community on using our catalogue of WDs, we developed a web archive system that can be accessed from a webpage\footnote{\url{http://svocats.cab.inta-csic.es/wdw/index.php}} or through a VO ConeSearch. \footnote{For example, \url{http://svocats.cab.inta-csic.es/wdw7/cs.php?RA=301.08&DEC=-67.482&SR=1&VERB=2}}

The archive system implements a very simple search interface that permits queries by coordinates and radius as well as by other parameters of interest. The user can also select the maximum number of sources (with values from 10 to unlimited) and the number of columns to return (minimum, default, or maximum verbosity).

The result of the query is a HTML table with all the sources found in the archive fulfilling the search criteria. The result can also be downloaded as a VOTable or a CSV file. Detailed information on the output fields can be obtained placing the mouse over the question mark (‘?’) located close to the name of the column. The archive also implements the Simple Application Messaging Protocol\footnote{\url{ http://www.ivoa.net/documents/SAMP}} (SAMP) VO protocol. SAMP allows VO applications to communicate with each other in a seamless and transparent manner for the user. This way, the results of a query can be easily transferred to other VO applications, such as, e.g. \texttt{TOPCAT}.

\begin{table*}[ht!]
\caption[]{Description of the columns in the catalogue of white dwarfs with reliable and tentative infrared excess.}
\label{tab:Cat}
\centering
\begin{tabular}{lll}
\noalign{\smallskip}
\hline
\hline
\noalign{\smallskip}
Column name                      & Units           & Description                                                              \\
\noalign{\smallskip}
\hline
\noalign{\smallskip}
\G{}\_id                         &                 & \G{} DR3 source identifier                                               \\
RA                               & deg             & \G{} DR3 right ascension (Epoch J2016)                                   \\
DEC                              & deg             & \G{} DR3 declination (Epoch J2016)                                       \\
parallax                         & mas             & \G{} DR3 parallax                                                        \\
Distance                         & pc              & Distance calculated using the \G{} parallax                              \\
$G_{abs}$                        & mag             & \G{} absolute G-band mean magnitude                                      \\
G                                & mag             & \G{} G-band magnitude                                                    \\
$G_{bp}$                         & mag             & \G{} $G_{BP}$ magnitude                                                  \\
$G_{rp}$                         & mag             & \G{} $G_{RP}$ magnitude                                                  \\
Excess\_type                     &                 & Type of the excess: reliable or tentative                                \\
Excess\_beginning                &                 & Band where the excess beginning                                          \\
Ref.                             &                 & Reference about the excess                                               \\    
PDA\_JE+23                       &                 & Probability of being DA (>0.5) or non-DA(<0.5) of JE+23                  \\
$T_{eff}$\_JE+23                 & K               & Effective temperature of JE+23                                           \\
$\log{g}$\_JE+23                 & log(cm/s$^2$)   & Surface gravity of JE+23                                                 \\
M\_JE+23                         & $M_{\odot}$     & Mass of JE+23                                                            \\
2MASS\_id                        &                 & 2MASS source identifier                                                  \\
UKIDSS\_cat                      &                 & UKIDSS catalogue                                                         \\
UKIDSS\_id                       &                 & UKIDSS source identifier                                                 \\
VISTA\_cat                       &                 & VISTA catalogue                                                          \\
VISTA\_id                        &                 & VISTA source identifier                                                  \\
WISE\_id                         &                 & WISE source identifier                                                   \\
Spitzer\_cat                     &                 & Spitzer catalogue                                                        \\
Spitzer\_id                      &                 & Spitzer source identifier                                                \\
SED\_phot.                       &                 & Link to the source SED photometry                                        \\
SED\_graphic                     &                 & Link to the source SED graphic                                           \\
\noalign{\smallskip}
\hline
\noalign{\smallskip}
\end{tabular}
\end{table*}

\newpage
\section{Spectroscopically confirmed infrared excess table}

\begin{table*}[ht!]
	\caption[]{Spectroscopically confirmed infrared excess.}
	\label{tab:spec_confirmed}
	\centering
    \small
    	\begin{tabular}{r c c c c}
    	\hline
    	\hline
    	\noalign{\smallskip}
    	\G{} DR3 id & This work excess class. & Ref. & Ref. system class. & Ref. telescope (instrument)\\
    	\hline
    	\noalign{\smallskip}
    	291057843317534464 & reliable & \begin{tabular}{c} {[1]} \\ {[2]} \end{tabular} & \begin{tabular}{c} WD + disk \\ WD + disk \end{tabular} & \begin{tabular}{c}         {Keck (MOSFIRE)} \\ {JWST (MIRI and NIRSpec)} \end{tabular} \\
    	\noalign{\smallskip}
        325899163483416704 & reliable & [3] & WD+ disk & KECK (NIRES) \\
        \noalign{\smallskip}
    	962995581174346112 & reliable & [1] & WD + BD & Keck (MOSFIRE) and Gemini (GNIRS) \\
        \noalign{\smallskip}
    	1081504483467714176 & reliable & {[4]} & WD + disk & JWST (MIRI) \\
        \noalign{\smallskip}
        1336442472164656000 & reliable & \begin{tabular}{c} {[5]} \\ {[6]} \\ {[7]}  \\ {[8]} \end{tabular} & \begin{tabular}{c} {WD + disk} \\ {WD + disk} \\ {WD + disk}  \\ {WD + disk} \end{tabular}  & \begin{tabular}{c} {IRTF (SpeX)} \\ {IRTF (SpeX)} \\ {Spitzer (IRS)}  \\ {IRTF (SpeX)} \end{tabular}  \\
        \noalign{\smallskip}
    	1429618420396285952 & reliable & [4] & WD + disk & JWST (MIRI) \\
        \noalign{\smallskip}
    	1641326979142898048 & reliable & \begin{tabular}{c} {[9]} \\ {[10]} \\ {[4]} \end{tabular} & \begin{tabular}{c} {WD + disk} \\ {WD + disk} \\ {WD + disk} \end{tabular} & \begin{tabular}{c} {IRTF(SpeX) } \\ {IRTF(SpeX) } \\ {JWST (MIRI)} \end{tabular} \\
        \noalign{\smallskip}
    	1837948790953103232 & reliable & [1] & WD + disk & Keck (MOSFIRE) \\
        \noalign{\smallskip}
    	2208124536065383424 & reliable & [11] & WD + companion & WHS (ISIS) \\
        \noalign{\smallskip}
        2307289214897332480 & reliable & [12] & WD + BD & VLT (X-Shooter) \\
        \noalign{\smallskip}
    	2416481783371550976 & reliable & {[13]} & WD + BD & VLT (X-Shooter) \\ 
        \noalign{\smallskip}
    	2588874825669925504 & reliable & \begin{tabular}{c} {[14]} \\ {[15]} \\{[16]} \end{tabular} & \begin{tabular}{c} WD + BD \\ WD + BD \\ WD + BD  \end{tabular} & \begin{tabular}{c} {VLT (X-Shooter) and HET (HRS)} \\ {VLT (X-Shooter)} \\ {IRTF (SpeX)} \end{tabular} \\ 
        \noalign{\smallskip}
    	2660358032257156736 & reliable & \begin{tabular}{c} {[17]} \\ {[6]} \\ {[8]} \\ {[18]} \\ {[10]}  \end{tabular} & \begin{tabular}{c} {WD + companion} \\ {WD + disk} \\ {WD + disk} \\ {WD + disk} \\ {WD + disk} \end{tabular} &\begin{tabular}{c} {IRTF (CGAS)} \\ {IRTF (SpeX)} \\ {IRTF (SpeX)} \\ {Spitzer (IRS) \& IRTF (SpeX)} \\ {IRTF (SpeX)}  \end{tabular} \\
        \noalign{\smallskip}
        3251748915515143296 & reliable & \begin{tabular}{c} {[6]} \\ {[8]} \end{tabular} & \begin{tabular}{c} {WD + disk} \\ {WD + disk} \end{tabular} & \begin{tabular}{c} {IRTF (SpeX)} \\ {IRTF (SpeX)} \end{tabular} \\
        \noalign{\smallskip}        
        3571559292842744960 & reliable & \begin{tabular}{c} {[8]} \\ {[19]} \end{tabular} & \begin{tabular}{c} {WD + disk} \\ {WD + disk} \end{tabular} & \begin{tabular}{c} {IRTF (SpeX)} \\ {IRTF (SpeX) \& UKIRT (UIST)} \end{tabular}\\
        \noalign{\smallskip} 
        3810933247769901696 & reliable & \begin{tabular}{c} {[6]} \\ {[10]} \end{tabular} & \begin{tabular}{c} {WD + disk} \\ {WD + disk} \end{tabular} & \begin{tabular}{c} {IRTF (SpeX)} \\ {IRTF (SpeX)} \end{tabular} \\
        \noalign{\smallskip}         
        3888723386196630784 & reliable & \begin{tabular}{c} {[6]} \\ {[10]} \end{tabular} & \begin{tabular}{c} {WD + disk} \\ {WD + disk}  \end{tabular} & \begin{tabular}{c} {IRTF (SpeX)} \\ {IRTF (SpeX)} \end{tabular} \\
        \noalign{\smallskip}
        4287654959563143168 & tentative & \begin{tabular}{c} {[20]} \\ {[10]} \end{tabular} & \begin{tabular}{c} {WD + disk} \\ {WD + disk} \end{tabular} & \begin{tabular}{c} {Magellan (FIRE)} \\ {IRTF (SpeX)} \end{tabular} \\
        \noalign{\smallskip}
    	4795556287084999552 & reliable & [4] & WD + disk & JWST (MIRI) \\
    	  \noalign{\smallskip}
    	5135466183642594304 & reliable & \begin{tabular}{c} {[21]} \\ {[22]} \end{tabular} & \begin{tabular}{c} WD + BD\\ WD + BD \end{tabular}  & \begin{tabular}{c} {Keck (NIRSPEC)} \\ {Keck (NIRSPEC)} \end{tabular} \\
    	\noalign{\smallskip}
        5261048587838041856 & reliable & [4] & WD + disk & JWST (MIRI) \\
    	\hline
        \noalign{\smallskip}
        948246835277407744  & no excess & [4] & WD + disk & JWST (MIRI) \\
        \noalign{\smallskip}
        4646535078125821568 & no excess & [23]  & WD + disk/exoplanet & JWST (MIRI) \\
        \noalign{\smallskip}
        4794599638953519360 & no excess & [4] & WD + disk & JWST (MIRI) \\
        \noalign{\smallskip}
        5560329193290547072 & no excess & [4] & WD + disk & JWST (MIRI) \\
        \noalign{\smallskip}
        5622626147732121856 & no excess & [4] & WD + disk & JWST (MIRI) \\
        \noalign{\smallskip}
        \hline        
        \hline
        \end{tabular}
    	\tablefoot{\footnotesize References: [1] \cite{Owens+23}, [2] \cite{Swanetal2024}, [3] \cite{Debesetal2019}, [4] \cite{Farihietal2025}, [5] \cite{Kilicetal2005}, [6] \cite{Kilicetal2006}, [7] \cite{Juraetal2007b}, [8] \cite{Kilicetal2007}, [9] \cite{Kilicetal2012}, [10] \cite{Barberetal2012}, [11] \cite{UndaSanzanaetal2008}, [12] \cite{Neustroevetal2023}, [13] \cite{Rebassa-Mansergas+22}, [14] \cite{Steele+13}, [15] \cite{Longstaffetal2019}, [16] \cite{Casewelletal2020}, [17] \cite{Tokunaga1988}, [18] \cite{Reachetal2009}, [19] \cite{Kilicetal2008}, [20] \cite{Melisetal2011}, [21] \cite{Farihi&Chistopher04}, [22] \cite{Dobbie+05}, [23] \cite{Limbach+24}.}        
\end{table*}

\end{appendix}

\end{document}